\documentclass[sigconf]{acmart}
\AtBeginDocument{%
  }

\copyrightyear{2026}
\acmYear{2026}
\setcopyright{cc}
\setcctype{by}
\acmConference[SIGMOD Companion '26]{Companion of the International Conference on Management of Data}{May 31-June 05, 2026}{Bengaluru, India}
\acmBooktitle{Companion of the International Conference on Management of Data (SIGMOD Companion '26), May 31-June 05, 2026, Bengaluru, India}
\acmDOI{10.1145/3788853.3801612}
\acmISBN{979-8-4007-2450-3/2026/05}

\usepackage{enumitem}

\usepackage{xspace}

\newcommand{\sys}{\textsc{DeepEye}\xspace}

\newcommand{\stab}{\vspace{1.2ex}\noindent}
\newcommand{\stitle}[1]{\stab\noindent{\textbf{#1}}}
\newcommand{\etitle}[1]{\vspace{1mm}\noindent{\underline{\em #1}}}

\begin{document}

\title{DeepEye: A Steerable Self-driving Data Agent System}

\author{Boyan Li}
\affiliation{%
  \institution{The Hong Kong University of Science and Technology (Guangzhou)}
  \city{Guangzhou}
  \country{China}}
\email{bli303@connect.hkust-gz.edu.cn}

\author{Yiran Peng}
\affiliation{%
  \institution{The Hong Kong University of Science and Technology (Guangzhou)}
  \city{Guangzhou}
  \country{China}}
\email{yiranpeng@hkust-gz.edu.cn}

\author{Yupeng Xie}
\affiliation{%
  \institution{The Hong Kong University of Science and Technology (Guangzhou)}
  \city{Guangzhou}
  \country{China}}
\email{yxie740@connect.hkust-gz.edu.cn}

\author{Sirong Lu}
\affiliation{%
  \institution{The Hong Kong University of Science and Technology (Guangzhou)}
  \city{Guangzhou}
  \country{China}}
\email{slu075@connect.hkust-gz.edu.cn}

\author{Yizhang Zhu}
\affiliation{%
  \institution{The Hong Kong University of Science and Technology (Guangzhou)}
  \city{Guangzhou}
  \country{China}}
\email{yzhu305@connect.hkust-gz.edu.cn}

\author{Xing Mu}
\affiliation{%
  \institution{The Hong Kong University of Science and Technology (Guangzhou)}
  \city{Guangzhou}
  \country{China}}
\email{xmu159@connect.hkust-gz.edu.cn}

\author{Xinyu Liu}
\affiliation{%
  \institution{The Hong Kong University of Science and Technology (Guangzhou)}
  \city{Guangzhou}
  \country{China}}
\email{xliu371@connect.hkust-gz.edu.cn}

\author{Yuyu Luo}
\affiliation{%
  \institution{The Hong Kong University of Science and Technology (Guangzhou)}
  \city{Guangzhou}
  \country{China}}
\email{yuyuluo@hkust-gz.edu.cn}

\renewcommand{\shortauthors}{Boyan Li et al.}

\begin{abstract}
% Large Language Models (LLMs) have revolutionized natural language interaction with data, but existing ``ChatBI'' systems still face challenges in performing joint analysis across heterogeneous data (databases, data files, documents) and are prone to ``context explosion'' in complex and iterative data tasks.
% Large Language Models (LLMs) have revolutionized natural language interaction with data. While the ultimate goal is to build autonomous ``Data Agents'', existing implementations are mostly limited to linear ``ChatBI'' systems.
% These systems still face challenges in performing joint analysis across heterogeneous data (databases, data files, documents) and are prone to ``context explosion'' in complex and iterative data tasks.
Large Language Models (LLMs) have revolutionized natural language interaction with data. The ``\textit{holy grail}'' of data analytics is to build autonomous \textit{Data Agents} that can self-drive complex data analysis workflows.
However, current implementations are still limited to linear ``ChatBI'' systems. These systems struggle with joint analysis across heterogeneous data sources (e.g., databases, documents, and data files) and often encounter ``context explosion'' in complex and iterative data analysis workflows. To address these challenges, we present \textbf{\sys}, a production-ready data agent system that adopts a \textit{workflow-centric architecture} to ensure scalability and trustworthiness.
\sys introduces a \textbf{Unified Multimodal Orchestration} protocol, enabling seamless integration of structured and unstructured data sources.
To mitigate hallucinations, it employs \textbf{Hierarchical Reasoning} with context isolation, decomposing complex intents into autonomous \textit{AgentNodes} and deterministic \textit{ToolNodes}.
Furthermore, \sys incorporates a database-inspired \textbf{Workflow Engine} (comprising a Compiler, Validator, Optimizer, and Executor) that guarantees structural correctness and accelerates execution via runtime topological optimization.
In this demonstration, we showcase \sys's ability to orchestrate complex workflows to generate diverse multimodal outputs -- including \textit{Data Videos, Dashboards, and Analytical Reports} -- highlighting its advantages in transparent execution, automated optimization, and human-in-the-loop reliability.
\end{abstract}




\maketitle

%%%%%%%%%%%%%%%%% Main Body of Paper %%%%%%%%%%%%%%%%%
\begin{figure*}[t!]
    \centering
    % \vspace{-1em}
    \includegraphics[width=\textwidth]{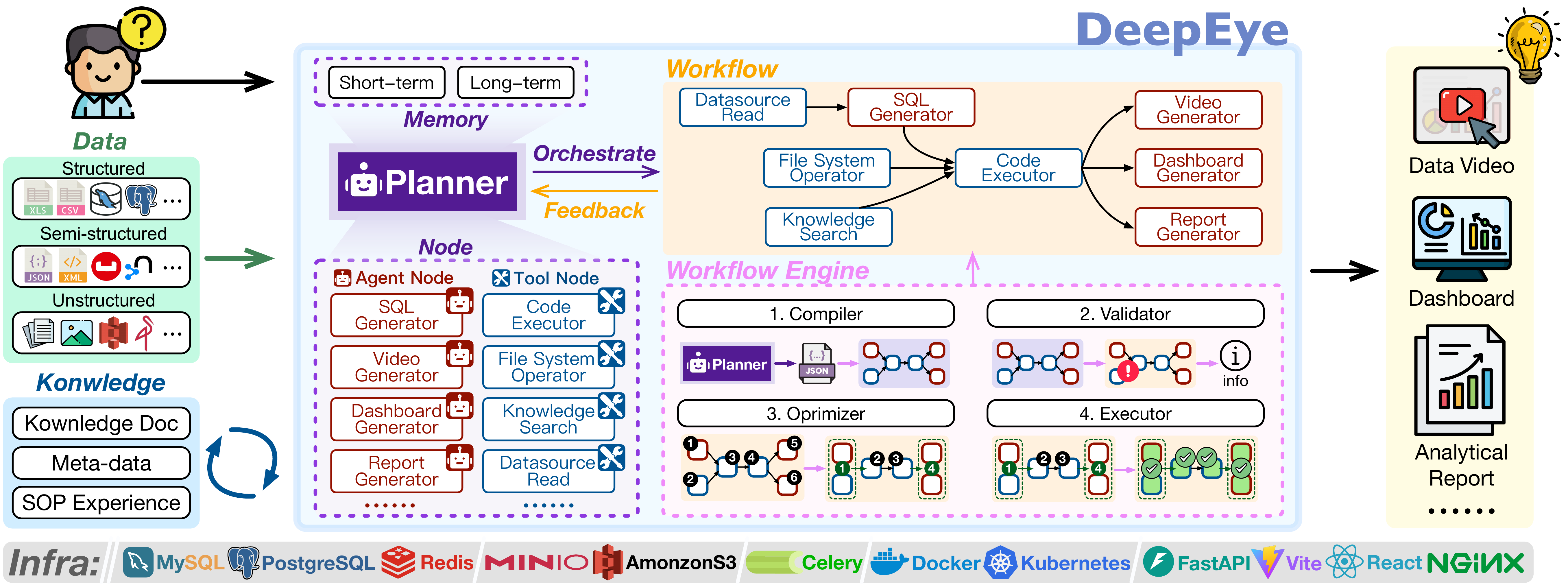}
    \vspace{-2em}
    \caption{The System Architecture of \sys.}
    \label{fig:overview}
    \vspace{-1em}
\end{figure*}

\section{Introduction}

The rapid evolution of Large Language Models (LLMs) has fundamentally transformed the landscape of data analytics, shifting the paradigm from rigid, menu-driven interfaces to flexible natural language interactions. 
Enterprises increasingly demand ``Data Agents'' capable of autonomously interpreting user intents, querying vast data repositories, and generating actionable insights~\cite{zhu2025survey}.
Unlike traditional Business Intelligence (BI) tools that require manual operation, these intelligent agents promise to automate the end-to-end analysis pipeline, democratizing access to data insights for non-technical users~\cite{luo2025nvbench}.

% \stitle{Prior Art.}
% Recent advancements have primarily focused on two distinct tracks: Text-to-SQL systems and general-purpose Agents. 
% Text-to-SQL approaches have achieved remarkable success in translating natural language into executable queries for structured databases~\cite{lialpha,li2025deepeye}. 
% Meanwhile, agent frameworks utilize LLMs to plan and execute sequences of tool calls~\cite{zhangaflow}. 
% However, deploying these systems in production-grade data environments remains elusive. 
% Most existing data agents operate as linear ``ChatBI'' bots, whose fragility and opacity hinder the development of robust, production-ready systems.
\stitle{Prior Art \& Limitations.} 
To realize this vision, recent advancements have primarily diverged into two streams: specialized \textit{Text-to-SQL} methods for structured databases~\cite{lialpha,li2025deepeye} and general-purpose \textit{Agent frameworks} that utilize LLMs for tool orchestration~\cite{zhangaflow}. 
However, deploying these systems in production environments remains elusive. 
Most existing solutions operate either as siloed query translators or linear ``ChatBI'' bots. 
Their inherent \textit{fragility} in handling complex, multi-step workflows and \textit{opacity} in execution logic hinder their adoption in trustworthy enterprise scenarios.

% \stitle{Challenges.}
% Despite the progress, building a robust, production-ready data agent system faces three critical challenges.
% \textbf{(C1)} \textit{The Heterogeneity Gap.} Real-world decision-making rarely relies on a single data modality. A comprehensive analysis often requires cross-referencing structured data (e.g., sales records in SQL) with unstructured knowledge (e.g., market reports in PDF). Current systems are mostly siloed: Text-to-SQL handles only databases, while RAG handles only documents, lacking unified frameworks capable of \textit{joint analysis} across heterogeneous sources.
% \textbf{(C2)} \textit{Context Explosion in Complex Reasoning.} Complex analytical tasks involve multi-step reasoning and diverse tool usage. Traditional \textit{Single-Agent architectures} force the descriptions of all available tools, intermediate observations, and reasoning history into a single global context window. This ``Context Explosion'' inevitably leads to LLM hallucinations, loss of focus, and failure to follow instructions as the workflow complexity increases.
% \textbf{(C3)} \textit{Unreliability and Inefficiency.} The non-deterministic nature of LLMs clashes with the requirement for rigorous data engineering. Existing agents often generate ``black-box'' execution chains that are difficult to validate, debug, or optimize. Furthermore, they often execute tasks sequentially, missing opportunities for parallelism (e.g., fetching data while summarizing text), leading to higher latency.
\stitle{Challenges.}
% 关键修改：用一句话把上一段的“fragility/opacity”链接到下面的C1-C3
We attribute these limitations to three fundamental challenges that current architectures fail to address:
\textbf{(C1)} \textit{The Heterogeneity Gap.} Real-world decision-making rarely relies on a single data modality. A comprehensive analysis often requires cross-referencing structured data (e.g., sales records in SQL) with unstructured knowledge (e.g., market reports in PDF). Current systems are mostly siloed: Text-to-SQL handles only databases, while RAG handles only documents, lacking unified frameworks capable of \textit{joint analysis} across heterogeneous sources.
\textbf{(C2)} \textit{Context Explosion in Complex Reasoning.} Complex analytical tasks involve multi-step reasoning and diverse tool usage. Traditional \textit{Single-Agent architectures} force the descriptions of all available tools, intermediate observations, and reasoning history into a single global context window. This ``Context Explosion'' inevitably leads to LLM hallucinations, loss of focus, and failure to follow instructions as the workflow complexity increases.
\textbf{(C3)} \textit{Unreliability and Inefficiency.} The non-deterministic nature of LLMs clashes with the requirement for rigorous data engineering. Existing agents often generate ``black-box'' execution chains that are difficult to validate, debug, or optimize. Furthermore, they often execute tasks sequentially, missing opportunities for parallelism (e.g., fetching data while summarizing text), leading to higher latency.

% \stitle{Our Approach: \sys.}
% To address these challenges, we present \sys, a \textit{steerable self-driving} data agent system. 
% \sys departs from the linear chat paradigm, adopting a \textit{workflow-centric architecture} that orchestrates data analysis as transparent Directed Acyclic Graphs (DAGs).
% \sys introduces three key innovations.
% First, to tackle \textbf{C1}, \sys features \textit{Unified Multimodal Orchestration} by implementing a unified node protocol that standardizes interactions across AgentNodes and ToolNodes, enabling seamless joint analysis of DB, File, and Knowledge bases, and synthesizing results into Data Videos, Dashboards, and Analytical Reports.
% Second, addressing \textbf{C2}, \sys employs \textit{Hierarchical Reasoning}. It decomposes global intents into context-isolated sub-agents, where the planner manages high-level dependencies while sub-agents handle local reasoning with private memory, effectively mitigating context overflow.
% Third, solving \textbf{C3}, \sys incorporates a \textit{Database-Inspired Workflow Engine} comprising a \textit{Compiler}, \textit{Validator}, \textit{Optimizer}, and \textit{Executor}. This engine performs static structural validation and runtime topological optimization (e.g., automatic parallelization), ensuring trustworthy and efficient execution.

\stitle{Our Approach: \sys.}
To address these challenges, we present \sys, a \textit{steerable} data agent system. 
Departing from linear chat paradigms, \sys adopts a \textit{workflow-centric architecture} that orchestrates data analysis as transparent Directed Acyclic Graphs (DAGs).
\sys introduces three key innovations.
First, tackling \textbf{C1}, \sys features \textit{Unified Multimodal Orchestration}. Its standardized node protocol bridges heterogeneous components (AgentNodes/ToolNodes), enabling seamless joint analysis of DB, File, and Knowledge Bases, and synthesizing results into Data Videos, Dashboards, and Analytical Reports.
Second, addressing \textbf{C2}, \sys employs \textit{Hierarchical Reasoning}. It decomposes intents into context-isolated sub-agents; the planner manages global dependencies while sub-agents handle local reasoning, effectively mitigating context overflow.
Third, solving \textbf{C3}, \sys integrates a \textit{Database-Inspired Workflow Engine}. Through static validation and runtime topological optimization (e.g., automatic parallelization), the engine ensures trustworthy and efficient execution.

\stitle{Demonstration Scenarios.}
% In this demo, we showcase \sys through a comprehensive ``Global Sales Performance Analysis'' scenario. 
% Attendees will experience: 
% (1) \textbf{Context-Aware Orchestration}, where users utilize explicit context binding (via ``@'' referencing) to trigger the generation of runtime-optimized workflows that query SQL databases and analyze documents in parallel; 
% (2) \textbf{Transparent Human-in-the-loop Intervention}, demonstrating how users can inspect internal node parameters (e.g., generated SQL) and leverage the Validator to prevent schema mismatches during manual refinement; 
% and (3) \textbf{Multimodal Synthesis}, culminating in the automated generation of interactive Dashboards, narrated Data Videos, and analytical Reports.
% In this demo, we showcase \sys through a comprehensive ``Global Sales Performance Analysis'' case, organized into two key scenarios.
% First, we present \textit{Automated Multimodal Analysis}, where attendees will observe how users utilize explicit context binding (via ``@'' referencing) to trigger runtime-optimized workflows that query databases and analyze documents in parallel, culminating in the generation of Data Videos, Dashboards and Analytical Reports.
% Second, we demonstrate \textit{Human-in-the-Loop Refinement}, highlighting the system's transparency by allowing users to inspect internal node parameters (e.g., generated SQL) and leverage the Validator to prevent schema mismatches during manual intervention.
% A demonstration video is available online\footnote{\url{https://drive.google.com/file/d/1Klfga4bn0phrARtBoIr32sOWcy_QzZXd/view?usp=sharing}}.
We showcase \sys via a ``Global Sales Performance Analysis'' case.
First, \textit{Automated Multimodal Analysis} demonstrates how explicit context binding (via ``@'') triggers parallel database and document processing to synthesize Data Videos, Dashboards, and Reports.
Second, \textit{Human-in-the-Loop Refinement} highlights transparency by allowing users to inspect node internals (e.g., generated SQL) and use the Validator to prevent schema mismatches during manual edits.
A demonstration video is available online\footnote{\url{https://drive.google.com/file/d/1TvPbqv-JBfcSceXjlE26_f1LTovVvt-K/view?usp=sharing}}.

\begin{figure*}[t!]
    \centering    \includegraphics[width=.99\textwidth]{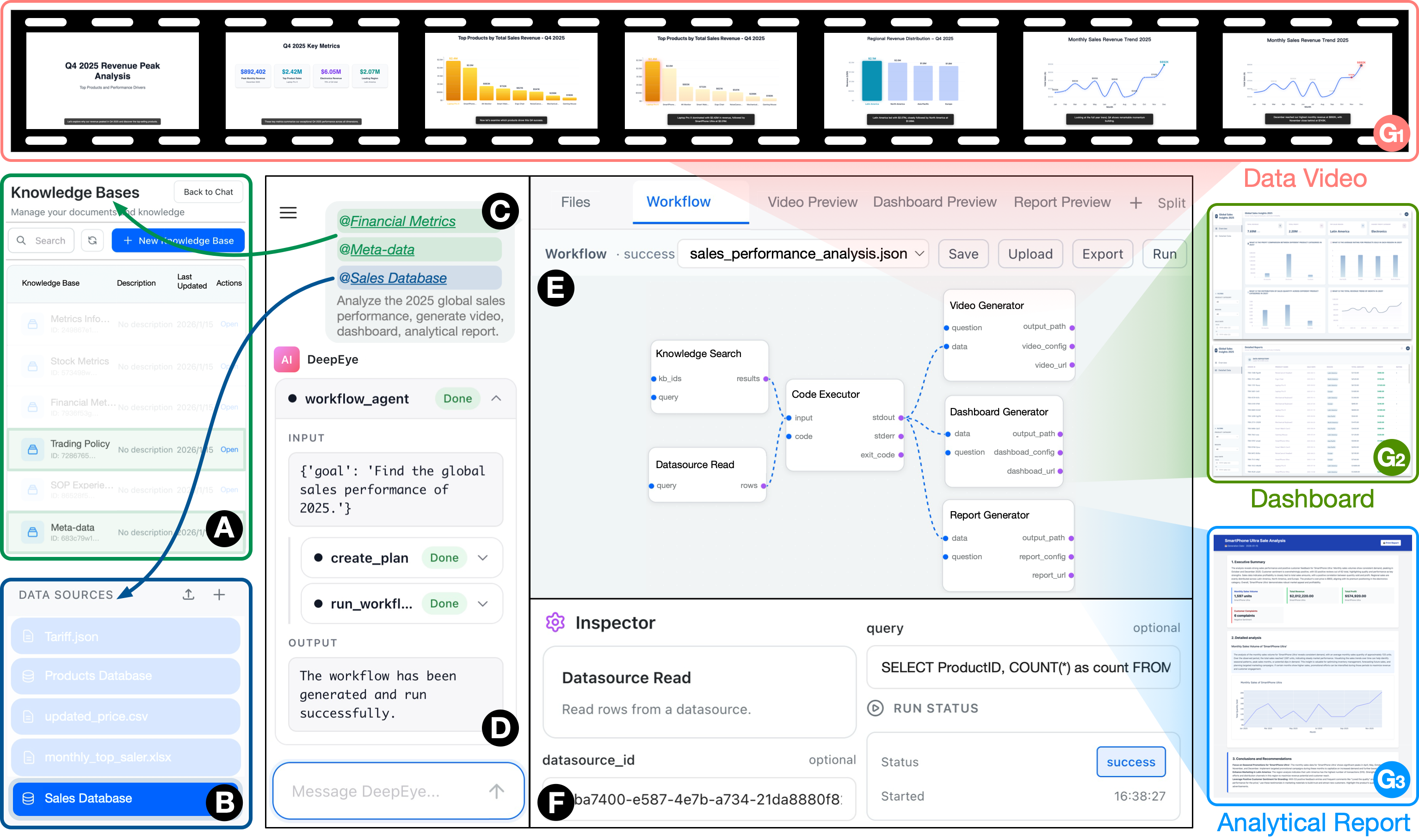}
    \vspace{-1em}
    \caption{The Demonstration of \sys.}
    \vspace{-1em}
    \label{fig:demo}
\end{figure*}

\section{System Architecture}

\sys implements a workflow-centric architecture that bridges the gap between flexible LLM reasoning and rigid data engineering. 
As shown in Figure~\ref{fig:overview}, the system is structured into four vertically integrated layers, underpinned by a cloud-native infrastructure.

\subsection{Unified Node Protocol and Isolation}
% To ensure extensibility and standard orchestration, \sys abstracts all analytical capabilities into a formal \textit{Unified Node Protocol}. 
To enable the \textit{Unified Multimodal Orchestration}, \sys abstracts all analytical capabilities into a formal \textbf{Unified Node Protocol}.
Formally, we define a generic node $\mathcal{N}$ as a 5-tuple:
{\setlength{\abovedisplayskip}{4pt}
\setlength{\belowdisplayskip}{4pt}
\begin{equation}
    \mathcal{N} = \langle \mathcal{D}, \mathcal{I}, \mathcal{O}, \mathcal{C}, \Phi \rangle
\end{equation}}
where:
\begin{itemize}[leftmargin=*]
    \item $\mathcal{D}$ denotes the \textit{Node Semantic Description}, a natural language string describing the node's functionality (e.g., \textit{``Generates a Python script to visualize data''}). The Planner relies on $\mathcal{D}$ for semantic retrieval and tool selection.
    \item $\mathcal{I} = \{p_1, \dots, p_m\}$ is the set of \textit{Input Ports}. Each port is strictly defined by a schema triplet $\langle \textit{Key}, \textit{Type}, \textit{Desc} \rangle$ (e.g., $\langle$``query'', \texttt{str}, ``The natural language question''$\rangle$). The description \textit{Desc} guides the Planner in accurately filling parameters from user context.
    \item $\mathcal{O} = \{q_1, \dots, q_n\}$ is the set of \textit{Output Ports}, similarly defined by schema and description, representing the standardized artifacts produced by the node.
    \item $\mathcal{C}$ denotes the \textit{Configuration Parameters} (e.g., model temperature), which are static environment settings.
    \item $\Phi: (\mathcal{I} \times \mathcal{C}) \to \mathcal{O}$ represents the internal execution logic.
\end{itemize}
% Based on the implementation of $\Phi$, we classify nodes into two categories:
Based on how $\Phi$ is implemented, we classify nodes into two types:

\stitle{ToolNodes (Deterministic Operators).} 
% These nodes encapsulate rule-based, atomic operations. The execution logic $\Phi_{tool}$ is a \textit{deterministic function} where the output strictly depends on the input: $\mathcal{O} = f_{code}(\mathcal{I}, \mathcal{C})$. 
% Examples include the \textit{DataConnector} and \textit{CodeExecutor}. 
% Since $\Phi_{tool}$ is devoid of probabilistic components, it guarantees idempotency and requires no reasoning context.
These nodes encapsulate rule-based atomic operations. 
Their execution logic $\Phi_{tool}$ is \textit{deterministic}, i.e., $\mathcal{O} = f_{code}(\mathcal{I}, \mathcal{C})$. 
Examples include \textit{DataConnector} and \textit{CodeExecutor}. 
Without probabilistic components, $\Phi_{tool}$ guarantees idempotency and requires no reasoning context.

\stitle{AgentNodes (Probabilistic Reasoners).} 
% These nodes act as autonomous sub-agents driven by LLMs. 
% The execution logic $\Phi_{agent}$ involves a probabilistic reasoning process or the execution of a nested sub-workflow (Sub-DAG). 
% Formally, $\Phi_{agent}$ utilizes a private, ephemeral context window $W_{local}$ and a set of internal tools $T_{internal}$ to derive the output:
These nodes are LLM-driven autonomous sub-agents. 
Their execution logic $\Phi_{agent}$ involves probabilistic reasoning or a nested sub-workflow (Sub-DAG). 
Formally, $\Phi_{agent}$ uses a private context window $W_{local}$ and internal tools $T_{internal}$ to derive:
{\setlength{\abovedisplayskip}{4pt}
\setlength{\belowdisplayskip}{4pt}
\begin{equation}
    \mathcal{O} \sim P(\cdot \mid \mathcal{I}, \mathcal{C}, W_{local}, T_{internal})
\end{equation}}
Crucially, \sys enforces \textit{Context Isolation}: the global planner only perceives the standardized I/O interfaces ($\mathcal{I}, \mathcal{O}$) and remains agnostic to $W_{local}$. 
% This encapsulation shields the global system from the noise of intermediate reasoning steps, effectively mitigating context overflow.
This encapsulation shields the global system from intermediate reasoning noise, mitigating context overflow.

%%%%%%%%%%%%%%%%%%%%%%%%%%%%%%%%%%%%%
%%%%%%%%%%%%%%%%%%%%%%%%%%%%%%%%%%%%%
%%%%%%%%%%%%%%%%%%%%%%%%%%%%%%%%%%%%%

\subsection{Memory-Augmented Planner}
\label{sec:planner}
% The \textit{Planner} serves as the orchestration brain, translating high-level user requests ($\mathcal{R}$) into executable workflow DAGs ($\mathcal{G}$). 
% It relies on a dual-memory architecture:
The \textit{Planner} translates high-level user requests ($\mathcal{R}$) into executable workflow DAGs ($\mathcal{G}$) using a dual-memory architecture:
\begin{itemize}[leftmargin=*]
    \item \textbf{Working Memory (Short-term)}: Maintains the \textit{Context Stack}, including multi-turn dialogue history, intermediate variable states, and runtime feedback from the engine.
    \item \textbf{Knowledge Base (Long-term)}: A vector database storing (1) Schema Metadata; (2) Domain Documentation; and (3) SOP Experience (verified historical workflows).
\end{itemize}

\stitle{Retrieval-Augmented Planning (Pre-Execution).} 
% Upon receiving $\mathcal{R}$, the Planner generates a query embedding to retrieve relevant SOPs and domain knowledge from the Knowledge Base. 
Upon receiving $\mathcal{R}$, the Planner retrieves relevant SOPs and domain knowledge from the Knowledge Base.
These retrieved contexts are injected as \textit{In-Context Demonstrations}. 
% Based on these few-shot examples, the Planner autonomously decides whether to strictly instantiate an existing template (for routine tasks) or synthesize a new topology (for novel tasks), ensuring high alignment with user intent.
The Planner either reuses an existing template or synthesizes a new topology, depending on the task.

\stitle{Self-Correction via Feedback (Runtime).} 
As shown in Figure~\ref{fig:overview}, the Planner forms a closed loop with the Workflow Engine. 
If execution fails (e.g., SQL syntax error), the error trace is fed back into the Working Memory. 
% The Planner triggers a \textit{Re-planning} process, analyzing the feedback to adjust node parameters or modify the DAG structure to resolve the issue dynamically.
The Planner then triggers \textit{Re-planning}, using the feedback to adjust node parameters or modify the DAG structure.

\stitle{Experience Accumulation (Post-Execution).} 
The system evolves over time. 
(1) \textit{Auto-Archiving}: Successfully executed workflows are automatically serialized and indexed into the Knowledge Base.
(2) \textit{Manual Creation}: Expert-refined workflows from the GUI canvas can be manually saved as high-quality SOP templates, allowing the system to learn from human expertise.

%%%%%%%%%%%%%%%%%%%%%%%%%%%%%%%%%%%%%
%%%%%%%%%%%%%%%%%%%%%%%%%%%%%%%%%%%%%
%%%%%%%%%%%%%%%%%%%%%%%%%%%%%%%%%%%%%

\subsection{The \sys Workflow Engine}
% At the core of the system lies the Workflow Engine, which transforms the Planner's logical plan into reliable, optimized execution. 
% Drawing inspiration from modern DBMS query engines, it processes workflows through four distinct phases:
At the core of the system, the Workflow Engine transforms the Planner's logical plan into reliable, optimized execution through four phases inspired by modern DBMS query engines:

\stitle{Phase 1: Compilation.} 
The \textit{Compiler} parses the logical JSON plan generated by the Planner into a structural \textit{DAG Object}. 
During this phase, it resolves variable references (e.g., linking the output artifact of a \textit{SQLNode} to the input port of a downstream \textit{DashboardNode}) and instantiates the concrete Node classes defined in the Protocol.

\stitle{Phase 2: Validation (Static Analysis).} 
To ensure safety before execution, the \textit{Validator} performs rigorous structural and semantic checks:
(1) \textit{Cycle Detection}: It employs Depth-First Search (DFS) to verify that the graph structure is acyclic.
(2) \textit{Schema Consistency}: It checks type compatibility between connected ports, ensuring that a downstream node receives data strictly matching its defined Input Schema $\langle Key, Type \rangle$.
(3) \textit{Completeness}: It verifies that all required secrets (e.g., DB credentials) are present in the secure vault.
% Any violation immediately triggers an error report to the user, preventing costly runtime failures.
Any violation is reported before execution, preventing runtime failures.

\stitle{Phase 3: Optimization (Runtime).} 
% Unlike simple linear chain agents, \sys incorporates a runtime \textit{Optimizer} to enhance efficiency.
% It analyzes the DAG topology using \textit{Kahn's Algorithm} to calculate the in-degree of each node.
Unlike linear-chain agents, \sys uses a runtime \textit{Optimizer} that analyzes the DAG with \textit{Kahn's Algorithm}.
Based on data dependencies, the optimizer groups nodes into \textit{Execution Layers}. 
Nodes within the same layer share no dependencies and are marked for \textit{Parallel Execution}. 
% For instance, a \textit{SQL Query} task and a \textit{PDF Summarization} task can be dispatched simultaneously to different workers, significantly reducing the end-to-end latency of complex multimodal tasks.
For instance, independent SQL and document tasks can be dispatched in parallel, reducing end-to-end latency.

\stitle{Phase 4: Execution.} 
The \textit{Executor} functions as the scheduler. 
It dispatches the optimized layers to the underlying asynchronous infrastructure (powered by Celery/Redis).
The Executor manages the data flow via a shared object store, handles retries for transient failures (e.g., API timeouts), and captures execution logs. 
% Critically, if a non-transient error occurs (e.g., execution exception), the detailed error trace is captured and fed back to the Planner, triggering the \textit{Self-Correction} mechanism described in Section~\ref{sec:planner}.
If a non-transient error occurs, the error trace is fed back to the Planner, triggering the \textit{Self-Correction} mechanism in Section~\ref{sec:planner}.

\subsection{Scalable Cloud-Native Infrastructure}
\sys uses a cloud-native stack for scalable deployment. Specifically, it adopts containerized microservices with \textit{Docker}, \textit{Docker Compose}, and a \textit{FastAPI} backend for consistent deployment and asynchronous request handling. To support long-running tasks without blocking the interface, the system uses \textit{Celery} and \textit{Redis} for asynchronous scheduling. It further combines \textit{PostgreSQL} for structured metadata and \textit{MinIO/S3} for unstructured artifacts through a unified storage interface. For security, ToolNode execution runs in \textit{sandboxed containers} with limited network access, while user data remains logically isolated.

\section{Demonstration}
% In this demonstration, we present \sys using a comprehensive ``Global Sales Performance Analysis'' scenario.
% The audience will experience the system's capabilities through the interactive interface shown in Figure~\ref{fig:demo}, highlighting the transition from heterogeneous data ingestion to multimodal insight generation.
We demonstrate \sys on a ``Global Sales Performance Analysis'' task through the interface in Figure~\ref{fig:demo}, covering the full pipeline from heterogeneous data ingestion to multimodal output generation.

\subsection{Scenario 1: Automated Multimodal Analysis}
\stitle{Goal.} To showcase the system's ability to autonomously orchestrate heterogeneous data sources and optimize workflow execution for diverse outputs.

\stitle{Workflow.}
The demonstration begins with data preparation. We upload sales records into the \textit{Data Sources Panel} (Fig. \ref{fig:demo}-B) and domain knowledge documents into the \textit{Knowledge Base} (Fig. \ref{fig:demo}-A).

\etitle{Intent Understanding with Context Binding:} In the \textit{Chat Panel} (Fig. \ref{fig:demo}-C), the user utilizes the system's ``@'' referencing feature to explicitly bind the analysis context. By selecting \textit{@Sales Database} and \textit{@Financial Metrics} (as shown in the figure), the user effectively grounds the subsequent natural language query: \textit{"Help me analyze the 2025 global sales performance."} This mechanism resolves ambiguity in data selection before planning begins.

\etitle{Orchestration \& Optimization:} The Planner analyzes the request and the bound contexts, as shown in Fig.~\ref{fig:demo}-D. It generates a logical plan which the Workflow Engine compiles into a DAG, visualized in the \textit{Workflow Canvas} (Fig. \ref{fig:demo}-E). 
\textit{Highlight:} We will draw attention to the \textit{Runtime Optimizer}. The audience will observe that the generated workflow branches out: a \textit{Knowledge Search Node} (for metrics) and a \textit{Datasource Read Node} (for the database) are arranged in the same execution layer. This visually demonstrates the engine's ability to identify independent tasks and execute them in parallel.

\etitle{Multimodal Synthesis:} 
%Upon successful execution, \sys synthesizes the results into three distinct formats shown in Fig. \ref{fig:demo}-G: a \textit{Data Video} (G1) narrating the insights, an interactive \textit{Dashboard} (G2) for data exploration, and a comprehensive \textit{Analytical Report} (G3). This confirms the system's capability to bridge the gap between structured data querying and unstructured content generation.
Upon successful execution, \sys synthesizes the results into a data video, dashboard, and analytical report (Fig.~\ref{fig:demo}-G), effectively bridging structured querying and multimodal content generation.

\subsection{Scenario 2: Human-in-the-Loop Refinement}
\stitle{Goal.} To demonstrate the transparency, editability, and trustworthiness of the workflow-centric architecture.

\stitle{Workflow.}
Unlike ``black-box'' ChatBI tools, \sys allows users to inspect and intervene at any stage.

\etitle{Transparent Inspection:} The user clicks on the \textit{SQL Generator Node} in the canvas. The \textit{Inspector Panel} (Fig. \ref{fig:demo}-F) opens, revealing the node's internal state. 
%
%Here, the user can inspect the strictly defined \textit{Input/Output Schema} and the specific \textit{node parameters} (e.g., the generated SQL query and configuration settings).
Here, the user can inspect the node’s \textit{I/O schema} and \textit{parameters}.
We integrated our Text-to-SQL technique, DeepEye-SQL~\cite{li2025deepeye},  to ensure high-accuracy query generation.

\etitle{Interactive Debugging:}
% We simulate a user intervention where they manually modify the SQL query logic within the Inspector.
We then simulate a manual SQL edit in the Inspector.
% \textit{Highlight:} When the user attempts to connect this modified node to an incompatible downstream node, the \textit{Workflow Validator} immediately flags a ``Schema Mismatch'' error on the canvas, preventing runtime failure.
When the edited node is connected to an incompatible downstream node, the \textit{Validator} immediately raises a ``Schema Mismatch'' error.

\etitle{Execution Feedback:}
Finally, we re-run the refined workflow. The \textit{Process Monitor} (Fig. \ref{fig:demo}-D) displays real-time logs and ``Thought-Action'' traces of the Planner, proving that the reasoning process is fully transparent and auditable.
% We finally re-run the workflow, and the Process Monitor (Fig.~\ref{fig:demo}-D) shows real-time logs and Planner traces for transparent debugging.
%%%%%%%%%%%%%%%%%%%%%%%%%%%%%%%%%%%%%%%%%%%%%%%%%%%%%%

\begin{acks}
This work was supported by the NSF of China (Grant No. 62402409).
\end{acks}

\bibliographystyle{ACM-Reference-Format}
\balance
\bibliography{main}

\appendix

\end{document}